\documentclass[12pt,preprint]{aastex}

\def\specchar#1{{\sc #1}}
\def\FeI{\mbox{Fe\,\specchar{i}}}

\def\CaII{\mbox{Ca\,\specchar{ii}}}

\shorttitle{Heating of magnetized chromosphere}

\shortauthors{Khomenko \& Collados}

\begin{document}

\title{Heating of the magnetized solar chromosphere by partial ionization effects}

\author{E. Khomenko\altaffilmark{1,2} and M. Collados\altaffilmark{1,2}}
\email{khomenko@iac.es}
\altaffiltext{1}{Instituto de Astrof\'{\i}sica de Canarias, 38205,
C/ V\'{\i}a L{\'a}ctea, s/n, La Laguna, Tenerife, Spain}
\altaffiltext{2}{Departamento de Astrof\'{\i}sica, Universidad de
La Laguna, 38205, La Laguna, Tenerife, Spain}

\begin{abstract}
In this paper, we study the heating of the magnetized solar chromosphere induced by the large fraction of neutral atoms present in this layer. The presence of neutrals, together with the decrease with height of the collisional coupling, leads to deviations from the classical MHD behavior of the chromospheric plasma. A relative net motion appears between the neutral and ionized components, usually referred to as ambipolar diffusion. The dissipation of currents in the chromosphere is enhanced orders of magnitude due to the action of ambipolar diffusion, as compared to the standard ohmic diffusion. We propose that a significant amount of magnetic energy can be released to the chromosphere just by existing force-free 10--40 G magnetic fields there. As a consequence, we conclude that ambipolar diffusion is an important process that should be included in chromospheric heating models, as it has the potential to rapidly heat the chromosphere. We perform analytical estimations and numerical simulations to prove this idea.
\end{abstract}


\section{Introduction}

The lower solar atmosphere $-$ photosphere and chromosphere $-$ is usually treated using the approximation of magnetohydrodynamics (MHD). Different dynamical processes, like, e.g., wave propagation, formation of magnetic structures, flux emergence, etc., have been quite successfully described with this formalism. Nevertheless, because of the rather cool temperatures of the photosphere and low chromosphere, the degree of ionization there is very small, reaching values as low as $10^{-4}$ in the temperature minimum, and remaining always well below unity even at larger heights \citep[see, e.g.][]{valc}. This fact, together with the decrease of collisional coupling with height, leads to a break of the assumptions underlying MHD.  In the upper photosphere and chromosphere, collisions are unable to couple completely the neutral and ionized components. Due to that, several new effects
appear, such as non-ideal Hall effect and ambipolar diffusion.

In plasma physics, the term ``ambipolar diffusion'' refers to the
diffusion of positive and negative particles at the same rate due
to their Coulomb interaction, which maintains the charge
neutrality at scales larger than the Debye length. In
astrophysics, however, ambipolar diffusion usually refers to the
decoupling of neutral and charged components. Ambipolar diffusion
causes the magnetic field to diffuse though neutral gas due to
collisions between neutrals and charged particles, the latter being
frozen-in into the magnetic field. The use of the same terminology
applied to these two different phenomena may lead to confusion. In
this paper we follow the astrophysical definition of the ambipolar
diffusion\footnote{Although the term ``ambipolar diffusion'' is
widely used in the literature, we consider more appropriate the
term ``diffusion by neutrals'', or ``neutral diffusion''. In the
rest of the paper we will use both terminologies
interchangeably.}.
As for the Hall effect, it appears as a result of the different
drift velocities of electrons and ions, which are not equally
affected by the presence of neutrals. The spatiotemporal scale
over which the Hall effect operates in a partially ionized plasma
is very different from the case of a fully ionized plasma
\citep{Pandey+Wardle2008}.

Recently, there has been an increasing number of works in the
literature showing the importance of the deviations from MHD in
different situations. The presence of neutral atoms in partially
ionized plasmas significantly affects wave excitation and
propagation \citep{Kumar+Roberts2003, Khodachenko2004,
Khodachenko2006, Forteza2007, Pandey2008, Vranjes2008, Soler2009,
Soler2010,Zaqarashvili2011}. The relative motion between the
neutral and charged species causes an increase of the collisional
damping of MHD waves in the photosphere, chromosphere and
prominence plasmas. This can be an important way to leak energy
from Alfv\'en and fast mode waves, causing the damping of coronal
loop oscillations \citep{Khodachenko2004, Khodachenko2006} and
prominence oscillations \citep{Forteza2007}. Under certain
conditions, the leakage of Alfv\'en wave energy can become the
source of instabilities in the medium due to the non-ideal Hall
effect \citep{Pandey2008}. The excitation rate of Alfv\'en waves
by foot-point motions of magnetic structures significantly
decreases if the neutral component is taken into account
\citep{Vranjes2008}.
%
Another effect is the possible appearance of cut-off
frequencies for Alfv\'en and kink waves, as mentioned by \citet{Soler2009}
\citep[see, however,][]{Zaqarashvili2011}.

The plasma partial ionization is also important for magnetic
reconnection, for the same reason as for waves.
\citet{Zweibel1989} has shown that the reconnection rate depends
strongly on the collisional coupling between ionized and neutral
species and \citet{Brandenburg+Zweibel1994,
Brandenburg+Zweibel1995} concluded that, due to the action of
ambipolar diffusion, oppositely oriented magnetic field lines can
be brought sufficiently close to facilitate the reconnection. In a
series of papers, \citet{Sakai2006}, \citet{Smith+Sakai2008}, and
\citet{Sakai+Smith2009} applied a two-fluid approximation to study
the reconnection between two current loops. The two-fluid approach
allowed these authors to find different temperatures of the
ionized and neutral species, and give rise to proton heating and
jet-like phenomena that, possibly, can be the reason for footpoint
heating of coronal loops, explosive events and jets in the
transition region and sunspot penumbrae.

All these non-ideal plasma effects can also modify the equilibrium
balance of photospheric and chromospheric structures. An
interesting mechanism to concentrate kilo-Gauss magnetic flux
tubes at photospheric level, due to the action of the Hall
current, was proposed by \citet{Khodachenko+Zaitsev2002}. At the
chromosphere, the order of magnitude stronger dissipation of
currents perpendicular to the magnetic field, compared to that of
longitudinal currents, was found to facilitate the creation of
potential force-free field structures \citep{Arber2009}. In
prominences, the frictional force generated due to collisions
between neutrals and ions was found to play a role in supporting
their structure against the gravitational force
\citep{Gilbert2002}.

Another phenomenon affected by non-ideal plasma effects is
magnetic flux emergence \citep{Leake+Arber2006, Arber2007}. These
authors showed that the amount of emerged flux is greatly
increased by the presence of a diffusive layer of partially
ionized plasma. Additionally, including neutrals removes the
non-realistic lifting of low-temperature plasma into the lower
corona and also avoids the subsequent Rayleigh-Taylor instability.

All the examples given above indicate that incorporating non-ideal
plasma effects is essential and can lead to new insights into the
physics of many phenomena taking place in the solar photosphere
and chromosphere. Despite this increasing evidence, we are far
from a complete understanding of the influence of these effects.
The aim of the present paper is to investigate the consequences of
the Joule dissipation of electric currents into the heating of the
magnetized chromosphere. In the presence of neutrals, the
ambipolar (or neutral) diffusion is expected to be orders of
magnitude larger than the classical ohmic diffusion, leading to
Reynolds number around unity \citep[see, e.g.,][]{Khodachenko2004,
Arber2007}. Diffusion by neutrals creates, in addition, anisotropy
in the properties of the plasma since currents perpendicular to
the magnetic field dissipate much quicker \citep{Arber2009}. Due
to that, one might expect important plasma heating. This work was
inspired by the recent research by \citet{DePontieu1998},
\citet{Judge2008} and \citet{Krasnoselskikh2010}, pointing in the
direction that current dissipation, enhanced by the presence of
neutrals in a plasma not entirely coupled by collisions, can play
an important role in the energy balance of the chromosphere and
above. In the rest of the paper, we will show analytical
calculations and numerical simulations in different models (from
purely academic to more realistic) of the amount of heating of
chromospheric flux tubes due to the ambipolar diffusion effect.

\section{Equations for partially ionized plasma}

Depending on the temporal and spatial scales under study, several
approaches can be adopted to describe the multi-component solar
plasma. Without going into a kinetic description (impossible to
use in the photosphere and lower chromosphere because they are too
dense), a detailed approach would consist in describing the three
components $-$ neutrals, ions and electrons $-$ by a system of
separate coupled equations (three-fluid approximation). However,
when the collisional coupling is large, a less detailed two-fluid
(adding up equations for electrons and ions) or single-fluid
(adding up all three equations) approximations can be used. Here
we adopt the latter approach. The derivation of the conservation
equations for individual species in this case can be found in,
e.g., \citet{Braginskii1965} and \citet{Bittencourt}. For
simplicity, we consider a hydrogen plasma with three components:
hydrogen ions ($i$), neutral hydrogen ($n$) and electrons ($e$)
and assume elastic collisions and no chemical reactions in the
system. A more complex composition of the plasma, to include other
atoms, has no relevance for our study, since one can always define
an average neutral or ionized atom. The equations of conservation
of mass, momentum and energy of the three species can be written
as follows:

\begin{eqnarray}
\label{eq:continuity3}
\frac{\partial \rho_e}{\partial t} + \vec{\nabla}(\rho_e\vec{u}_e)
& = & 0
\\ \nonumber
\frac{\partial \rho_i}{\partial t} + \vec{\nabla}(\rho_i\vec{u}_i)
& = & 0 \\  \nonumber
\frac{\partial \rho_n}{\partial t} + \vec{\nabla}(\rho_n\vec{u}_n)
& = & 0
\end{eqnarray}
\begin{eqnarray}
\label{eq:momentum3}
\rho_e\frac{D\vec{u}_e}{Dt} & = & -en_e(\vec{E} +
\vec{u}_e\times\vec{B}) + \rho_e\vec{g} - \vec{\nabla}\hat{p}_e +
\vec{R}_e
\\  \nonumber
\rho_i\frac{D\vec{u}_i}{Dt} & = & en_i(\vec{E} +
\vec{u}_i\times\vec{B}) + \rho_i\vec{g} - \vec{\nabla}\hat{p}_i +
\vec{R}_i
\\  \nonumber
\rho_n\frac{D\vec{u}_n}{Dt} & = &  \rho_n\vec{g} -
\vec{\nabla}\hat{p}_n + \vec{R}_n
\end{eqnarray}
\begin{eqnarray}
\label{eq:energy3}
\frac{1}{(\gamma-1)}\frac{D p_e}{Dt} +
\frac{p_e(\vec{\nabla}\vec{u}_e)}{(\gamma-1)} +
(\hat{p}_e\vec{\nabla})\vec{u}_e + \vec{\nabla}\vec{q}_e  & = &  -{\vec u}_e\vec{R}_e \\
\nonumber
\frac{1}{(\gamma-1)}\frac{D p_i}{Dt}  +
\frac{p_i(\vec{\nabla}\vec{u}_i)}{(\gamma-1)} +
(\hat{p}_i\vec{\nabla})\vec{u}_i + \vec{\nabla}\vec{q}_i  & = & -{\vec u}_i\vec{R}_i \\
\nonumber
\frac{1}{(\gamma-1)}\frac{D p_n}{Dt}  +
\frac{p_n(\vec{\nabla}\vec{u}_n)}{(\gamma-1)} +
(\hat{p}_n\vec{\nabla})\vec{u}_n + \vec{\nabla}\vec{q}_n & = &
-\vec{u}_n\vec{R}_n
\end{eqnarray}
where all the notations are standard. Similar system of equation
is used in e.g., \citet{Goedbloed+Poedts2004} and
\citet{Zaqarashvili2011}.
The terms $R_{\alpha}$ ($\alpha=e,i,n$) for elastic collisions can
be expressed as:
\begin{eqnarray}
\vec{R}_e & = & -\rho_e[ \nu_{ei}(\vec{u}_e - \vec{u}_i) +
\nu_{en}(\vec{u}_e - \vec{u}_n) ]
\\ \nonumber
\vec{R}_i & = & -\rho_i[ \nu_{ie}(\vec{u}_i - \vec{u}_e)  +
\nu_{in}(\vec{u}_i - \vec{u}_n) ]  \\
\nonumber
\vec{R}_n & = & -\rho_n[ \nu_{ni}(\vec{u}_n - \vec{u}_i)  +
\nu_{ne}(\vec{u}_n - \vec{u}_e) ]
\end{eqnarray}
the sum of all three being zero. Here $\nu_{\alpha\beta}$ are
collisional frequencies. We take these frequencies from \citet{Spitzer1962},
for collisions between the neutrals and charged particles,
and from \citet{Braginskii1965}, for collisions between the charged
particles:
\begin{equation}
\nu_{in}=n_{n}\sqrt{\frac{8 k_B T}{\pi m_{in}}}\sigma_{in}
\label{eq:nu_in}
\end{equation}
\begin{equation}
\nu_{en}=n_{n}\sqrt{\frac{8 k_B T}{\pi m_{en}}}\sigma_{en}
\label{eq:nu_en}
\end{equation}
\begin{equation}
\nu_{ei}= \frac{n_e e^4 \Lambda}{3 m_e^2 \epsilon_0^2} \left(
\frac{m_e}{2\pi k_B T} \right)^{3/2} \label{eq:nu_ei}
\end{equation}
where $m_{in}=m_i m_n/(m_i + m_n)$ and $m_{en}=m_e m_n/(m_e +
m_n)$. The respective cross sections are
$\sigma_{in}=5\times10^{-19}$ m$^2$ and $\sigma_{en}=10^{-19}$
m$^2$. $\Lambda$ is the Coulomb logarithm:
\begin{equation}
\Lambda=23.4 - 1.15\log_{10}{n_e}+3.45\log_{10}{T}
\end{equation}
\noindent with $n_e$ expressed in cgs units and T in eV.

We further assume that the pressure tensor $\hat{p}$ can be approximated by the scalar pressure and that the heat flux $\vec{q}$ is zero. Following \citet{Bittencourt}, we added up the equations for the three species. Under no further approximations, this leads to a system of quasi-MHD equations of the form:

\begin{eqnarray} \label{eq:continuity}
\frac{\partial \rho}{\partial t} +
\vec{\nabla}\left(\rho\vec{u}\right) = 0
\end{eqnarray}
\begin{equation} \label{eq:momentum}
\rho\frac{D\vec{u}}{D t}  = \vec{J}\times\vec{B} + \rho\vec{g} - \vec{\nabla}p
\end{equation}
\begin{equation}
\frac{1}{(\gamma-1)}\frac{D p}{D t} +
\frac{\gamma}{(\gamma-1)}p\vec{\nabla}\vec{u}= \vec{J} [\vec{E} + \vec{u}\times{\vec{B}} ]
\end{equation}

\noindent
In these equations,the following definitions are used:
\begin{equation}
\rho  = \rho_n + \rho_i + \rho_e
\end{equation}
\begin{equation}
\vec{u}  = \frac{\rho_n\vec{u}_n + \rho_i\vec{u}_i + \rho_e\vec{u}_e}{\rho}
\end{equation}
\begin{equation}
\vec{J} = en_e(\vec{u}_i - \vec{u}_e)
\end{equation}
\begin{equation}
p = \sum_{\alpha=n,i,e}p_{\alpha} + \frac{1}{3}\sum_{\alpha=n,i,e}\rho_{\alpha}w_{\alpha}^2
\end{equation}

\noindent
where
\begin{equation}
\vec{w}_{\alpha}  =  \vec{u}_{\alpha} - \vec{u}\,, \,\,\, \alpha=n,i,e,
\end{equation}
\noindent
is the diffusion velocity, i.e., the difference between the center of mass velocity of the system, $\vec{u}$, and the velocity of the individual components, $\vec{u}_{\alpha}$. We further consider that the diffusion velocities are small and neglect the term containing $w_{\alpha}^2$ in the definition of the total scalar pressure.

\begin{figure}[t]
\center
\includegraphics[width=9cm]{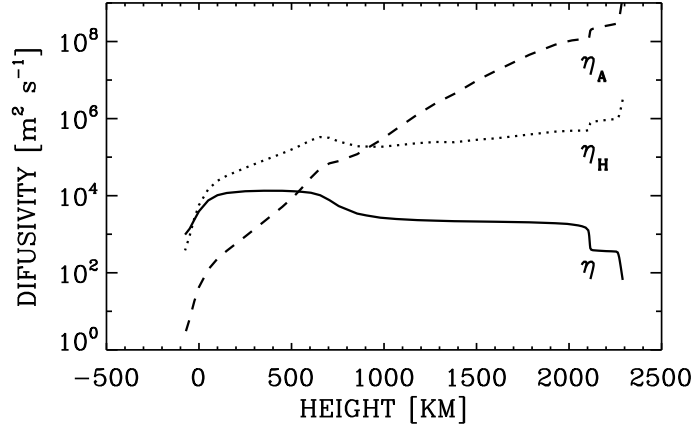}
\caption{{\footnotesize Diffusion coefficients from Eq. (\ref{eq:ohm}) calculated in the model atmosphere representing a 2nd-order thin magnetic flux tube \citep{Khomenko+etal2008, Pneuman+etal1986}. } }\label{fig:etas}
\end{figure}
\begin{figure}[t]
\center
\includegraphics[width=9cm]{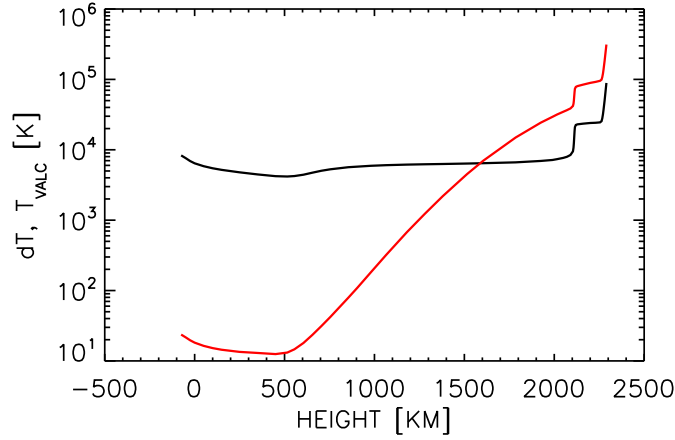}
\caption{{\footnotesize Temperature increment with height estimated after Eq. (\ref{eq:formula}) in the flux tube model atmosphere, assuming only 1\% of the existing magnetic field is dissipated (red line). Black line shows the temperature run of the VAL-C model atmosphere. } } \label{fig:dt}
\end{figure}

To close the system, the generalized Ohm's law is needed to describe the evolution of the currents. We use here the form of the Ohm's law from \citet{Braginskii1965} for slow processes:

\begin{equation}
\label{eq:ohm} \left[\vec{E} + \vec{u}\times\vec{B}\right]  = \eta\mu_0\vec{J} + \eta_H\frac{\mu_0}{|B|}\left[\vec{J}\times\vec{B}\right] + \eta_A\mu_0\vec{J_{\bot}}
\end{equation}

\noindent
where $\vec{J_{\bot}}$ is the component of current perpendicular to the magnetic field. This form of the generalized Ohm's law neglects the temporal
variations of the relative ion-neutral velocity, the effects on the currents by the partial pressure gradients of the three species, and the gravity force acting on electrons. The diffusion coefficients are given by the formulae:
\begin{eqnarray}
\eta & = &  \frac{m_e(\nu_{ei} + \nu_{en})}{e^2 n_e \mu_0} \label{eq:eta} \\
\eta_H & = & \frac{|B|}{en_e\mu_0} \label{eq:eta_H}\\
\eta_A & = & \frac{(\rho_n/\rho)^2|B|^2}{(\rho_i\nu_{in} + \rho_e\nu_{en})\mu_0} \label{eq:etaa}
\end{eqnarray}
The three terms on the right-hand side of the generalized Ohm's law are, in order, the classical ohmic term, the Hall term and the ambipolar or neutral term. The diffusion coefficients defined above depend on the collisional frequency, on the magnetic field strength and on the neutral fraction. It is important to note that, for a fully ionized plasma, the parameters $\rho_n$, $\nu_{in}$ and $\nu_{en}$ are zero. The squared exponent of $\rho_n$ makes $\eta_A$ vanish in this case, due to the linear dependence of the collisional frequencies with the density of neutrals (see Eqs.~\ref{eq:nu_in} and \ref{eq:nu_en}). To have an idea of the expected range of values of these coefficients, we have calculated them in a model atmosphere representing a 2nd-order thin flux tube \citep{Pneuman+etal1986, Khomenko+etal2008}. The field strength in this model decreases with height, from about 750 G in the photosphere to 37 G in the chromosphere, and the temperature structure is the same as VAL-C \citep{valc}. The results are given in Fig.~\ref{fig:etas}. The classical ohmic term reaches its largest values at photospheric heights, between 0 and 500 km. However, even there, the Hall term is about one order of magnitude larger than the ohmic term. The ambipolar term becomes dominant over the other two from 900 km
upwards. At chromospheric heights, this term is up to 5 orders of magnitude larger than the ohmic term.

Using the Ohm's law, the equation for internal energy takes the shape:
\begin{equation} \label{eq:intenergy}
\frac{1}{(\gamma - 1)}\frac{D p}{D t} + \frac{\gamma}{(\gamma - 1)} p\vec{\nabla}\vec{u} = \eta\mu_0\vec{J^2} + \eta_A\mu_0 \vec{J_{\bot}^2}
\end{equation}

As can be seen, the Hall term of the Ohm's law does not appear in the energy equation and, consequently, has no impact on the thermal evolution of the system. This is no surprise, given that $\eta_H$ does not depend on the collisions among the species (see Eq.~\ref{eq:eta_H}). Since we are interested in mechanisms that may lead to the heating of magnetic structures at chromospheric heights, we will concentrate in the rest of the paper on the action of the first and third terms of the Ohm's law and the Hall term will be removed from Eq.~\ref{eq:ohm}.

The induction equation for the temporal variations of the magnetic field is derived from the Ohm's law and Maxwell equations. It has the following form:
\begin{equation} \label{eq:induction}
\frac{\partial\vec{B}}{\partial t}  =  \vec{\nabla}\times \left[(\vec{u}\times\vec{B}) - \eta\mu_0\vec{J} - \eta_A\mu_0\vec{J_{\bot}} \right]
\end{equation}
Finally, combining the equations of momentum (Eq.~\ref{eq:momentum}), internal energy (Eq.~\ref{eq:intenergy}) and the induction equation (Eq.~\ref{eq:induction}), we obtain the following equation for the variations of the total energy:
\begin{eqnarray} \label{eq:totalenergy}
\frac{\partial e_{\rm tot}}{\partial t} + \vec{\nabla}\left(\vec{u}\left(e_{\rm tot} + p + \frac{B^2}{2\mu_0}\right) -  \frac{\vec{B}(\vec{u}\vec{B})}{\mu_0} \right) =  \\
\nonumber
\rho\vec{u}\vec{g}  +   \vec{\nabla}\left[\vec{B} \times (\eta + \eta_A) \vec{J}\right]
\end{eqnarray}
where $e_{\rm tot}=\rho u^2/2  + B^2/2\mu_0 + p/(\gamma-1)$.

Here, it is in order to note that the action of the ambipolar (neutral) term in the energy equation will stop in two cases. Firstly, when the plasma becomes fully ionized, since $\eta_A=0$. And, secondly, when the magnetic field evolves to a force-free configuration ($\vec{J}_{\bot}=0$),
since $\vec{J} \times \vec{B} =0$.

\section{Order of magnitude estimates}

The Joule dissipation of currents allows to transform the energy
stored in the magnetic field into thermal energy. Assuming a
scenario where the kinetic energy variations are negligible, we
can make an order of magnitude estimate of the amount of heating
that can be achieved by dissipating a given amount of magnetic
energy and fully converting it into thermal energy. One can write
that the loss of magnetic energy is equal to the increase of
thermal energy, which, to first order approximation, can be
expressed as
\begin{equation}
\frac{B_0\Delta B}{\mu_0} = \frac{\Delta p}{(\gamma-1)}.
\end{equation}
If we further assume that the perturbation of the density is
negligible and the totality of the pressure perturbations is
caused by the temperature variation (leading to the maximum
temperature increase by magnetic energy dissipation), we can write
$\Delta p = \rho R \Delta T$ and
\begin{equation}
\Delta T = \frac{(\gamma - 1)}{\rho R \mu_0}B_0 \Delta B
\label{eq:formula}
\end{equation}

This approximate equation gives us the amount of temperature increase, $\Delta T$, when the magnetic
field is decreased by an amount $\Delta B$. As an example of how large this energy conversion can
be, Fig.~\ref{fig:dt} gives the amount of the temperature increase $\Delta T$ according to Eq.
(\ref{eq:formula}), assuming that as little as 1\% of the total magnetic field at a given height is
dissipated. In this calculation, we used the same thermal and magnetic parameters of the flux tube
model as in Fig.~\ref{fig:etas}. We can compare the temperature increase $\Delta T$ (red curve) with
the actual temperature at each height given by the VAL-C atmospheric model (black curve). It shows
that an important temperature increase can be reached at heights above 1200--1300 km, already in the
lower chromosphere. Although the assumed 1\% is totally unmotivated, this example shows that a
very small conversion of magnetic energy, even below detectable limits in terms of field strength,
can lead to a significant amount of heating.

The time scales at which this dissipation happens depend on the
value of the ambipolar diffusion coefficient and on the typical
scales of the system. Assuming typical velocities of motion of
$V=10^4$ m s$^{-1}$, spatial scales of $L=10^5$ m and
$\eta_A=10^7-10^8$ (at 1500--2000 km, see Fig.~\ref{fig:etas}) the
order of magnitude estimate of the time scale gives:
\begin{equation}
t \approx L^2/\eta_A=100-1000\,\,\,\, {\rm sec.}
\end{equation}
This time scale is much shorter than in the case when only ohmic
diffusion is considered ($t \approx 10^7$ sec, i.e about 4
months). Thus, we may expect that important heating can be reached
due to ambipolar diffusion in a relatively short time interval of
the order of minutes.

\begin{figure}[t]
\center
\includegraphics[width=9cm]{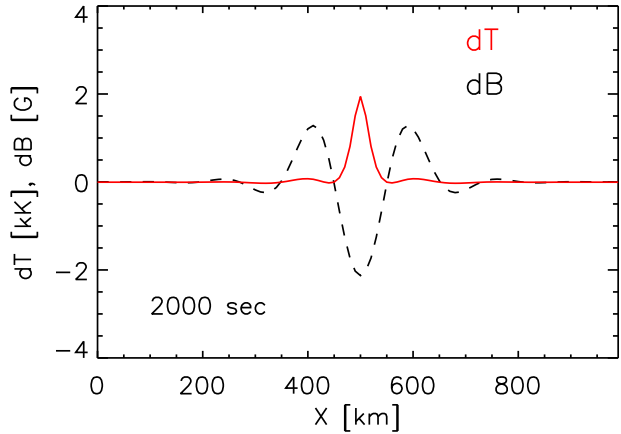}
\caption{{\footnotesize Variations of the magnetic field (black
dashed line) and temperature (red solid line) with respect to
their equilibrium values across the horizontal cut through the
vertical magnetic flux tube 2000 sec after the start of the
simulation.} } \label{fig:dtdb}
\end{figure}

\begin{figure}[t]
\center
\includegraphics[width=9cm]{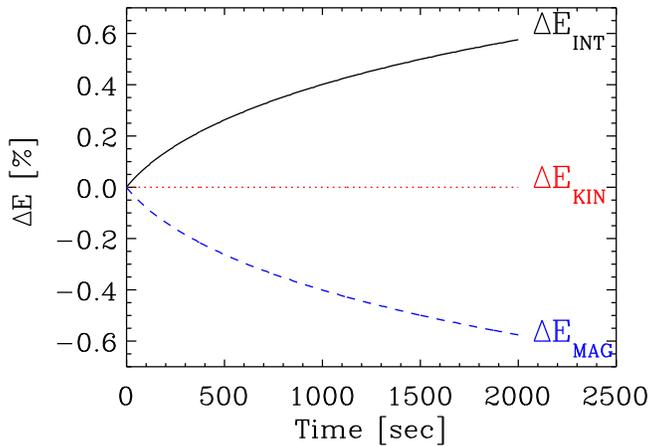}
\caption{{\footnotesize Time evolution of the variations of the internal energy ($\Delta E_{\rm INT}=E_{\rm INT}-E_{\rm INT}(0)$, black solid line);  magnetic energy ($\Delta E_{\rm MAG}=E_{\rm MAG}-E_{\rm MAG}(0)$, blue dashed line) and kinetic energy ($\Delta E_{\rm KIN}=E_{\rm KIN}-E_{\rm KIN}(0)$, red dotted line), in the simple unstratified model atmosphere containing a vertical flux tube. The energies are averaged over the horizontal direction. The variations are shown in per cents from the total (horizontally averaged) initial energy of the system $E_{\rm TOT}(0)=E_{\rm INT}(0)+E_{\rm MAG}(0)+E_{\rm KIN}(0)$.} } \label{fig:energyconst}
\end{figure}

The same applies to the Reynolds number. The Reynolds number defined
by the ambipolar diffusion is estimated as:
\begin{equation}
R_m \approx VL/\eta_A=10-100,
\end{equation}
while the Reynolds number for the ohmic diffusion in the same
conditions is much larger, $R_m \approx 10^6$.

\section{Unstratified atmosphere}

In this section, we consider the simple problem of an isothermal atmosphere without gravity. We assume the vertical magnetic field
depending only on horizontal coordinate as:
\begin{equation} \label{eq:vtube}
B_Z(x)=B_0\exp{\left(-\frac{(x-x_0)^2}{2\sigma^2} \right)}.
\label{eq:Bz}
\end{equation}
Gas pressures are prescribed to keep the model in magnetostatic equilibrium. The parameters $\sigma=40$ km and $B_0$ is defined later.

We solve numerically the equations of conservation of mass (Eq.~\ref{eq:continuity}), momentum (Eq.~\ref{eq:momentum}), total energy (Eq.~\ref{eq:totalenergy}), and the induction equation (Eq.~\ref{eq:induction}), after subtracting the equilibrium conditions from them. These equation are solved by means of our code {\sc mancha} \citep{Khomenko+etal2008b, Felipe+etal2010}. The code, as described in the above papers, solves the ideal MHD equations for non-linear perturbations to the magneto-static equilibrium, without physical diffusive terms. We modified the code to include the ohmic and ambipolar diffusion terms in the equation of energy conservation and in the induction equation, as described above. As our code solves the equations for non-linear perturbations, we treat the diffusion terms as perturbations. Without introducing this (or any other) perturbation no evolution is possible. The equations are solved in two spatial dimensions, though the vector quantities are allowed to have three dimensions (2.5D approximation). As the temperature, and the ionization state, varies with time, we recalculate the ionization balance of the atmosphere at each time step, assuming Local Thermodynamic Equilibrium (Saha equations). We then update the neutral fraction, $\rho_n/\rho$, needed for the calculation of the ambipolar diffusion coefficient (Eq.~\ref{eq:etaa}).

As initial atmospheric parameters far from the axis of the magnetic structure defined by Eq.~\ref{eq:Bz} , we take values of the temperature, pressure and density from VAL-C model at 1600 km. The value of the magnetic field at the center of the flux tube is limited by the condition of the horizontal pressure equilibrium ($P_{\rm mag}+P_{\rm gas}={\rm const}$). According to this condition, we set $B_0=18$ G.

\begin{figure}[t]
\center
\includegraphics[width=9cm]{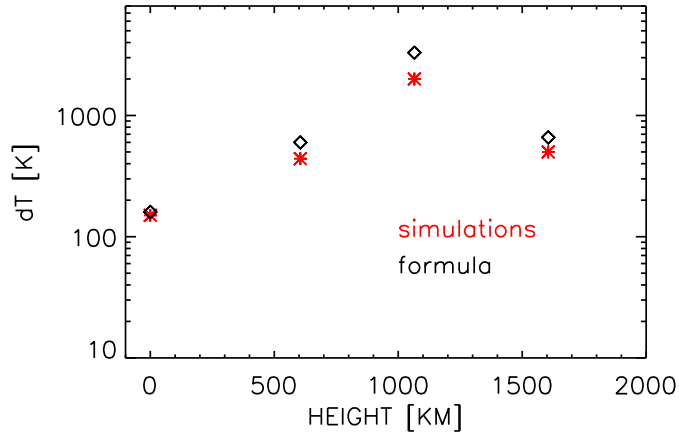}
\caption{{\footnotesize Temperature increase $\Delta T$ reached after the 2000 sec of the simulation time in the simple unstratified model atmosphere containing a vertical flux tube (red asterisk). The different points are for the four different simulations with thermal parameters taken from the VAL-C model atmosphere at the corresponding heights. The initial field strength is taken to be $B_0=1500$, 100, 18 and 3 G at 0, 600, 1000 and 1600 km, respectively. This field has been decreased by 9, 5, 2 and 0.2 G after the 2000 of the simulations. The black diamonds show the values of $\Delta T$ obtained from Eq.~\ref{eq:formula}.} } \label{fig:dtconst}
\end{figure}

The simple atmosphere considered in this section is initially in equilibrium, obtained without considering the diffusion terms. Without external perturbation, it does not evolve. But, after introducing the perturbation in the form of diffusion terms, we perturb the initial magnetic field structure via the induction equation. Then, as time evolves, this perturbation translates to the rest of the variables of the system and it
starts to change. Figure~\ref{fig:dtdb} shows the evolved system 2000 seconds after the introduction of the perturbation. At this time moment, the temperature at the center of the flux tube has increased by about 2000 K and the magnetic field has decreased by about 2 G.

The time variation of the different contributions to the total energy are given in Fig.~\ref{fig:energyconst}. It gives the internal, magnetic and kinetic energies, averaged over the horizontal direction of the simulation domain. The average kinetic energy is negligible compared to the other two and no important plasma motions are produced. The internal energy increases with time, and the magnetic energy decreases, in exactly the same
amount. Thus, in this simple experiment we convert all the released magnetic energy into internal energy by means of Joule dissipation of electric currents enhanced due to neutral diffusion (the contribution of the ohmic diffusion is very small). As no energy dissipation mechanisms are considered in this toy simulation, the internal energy (and temperature) would be expected to ever increase with time. As was noted at the end of Section 2, the action of the heating term would stop in two cases. In one possible scenario, once sufficiently high temperatures are reached, the plasma becomes totally ionized and the ambipolar diffusion coefficient becomes $\eta_A=0$. In the second scenario, the magnetic field evolves to a force-free configuration $\vec{J}_{\perp}=0$ and dissipates a small or large fraction of its energy, depending on the magnetic configuration. None of these two conditions are yet reached after 2000 sec of the simulation time shown in Fig.~\ref{fig:energyconst} and the internal energy shows a tendency to increase with time. 

Obviously, this calculation is just an illustration to show the efficiency of the proposed mechanism for chromospheric heating. In a more realistic situation some energy dissipation mechanism would balance the heating so that the temperature would not increase unbounded to ionize the whole chromosphere. The most probable mechanism to balance the heating comes from radiative energy losses, shown to be a key dissipation mechanism for the chromosphere since the work of \citet{Athay1966}.

We have repeated the above experiment for different thermodynamic and magnetic field parameters taken at different heights of the VAL-C model atmosphere. The resulting temperature increase reached after 2000 sec of the simulation time is shown in Fig.~\ref{fig:dtconst}. For comparison, we show the values of $\Delta T$ obtained from the simple formula (Eq.~\ref{eq:formula}). The most important heating is reached at 1000 km height. The values given by the formula are in order-of-magnitude agreement with those obtained in the numerical experiment, suggesting that this formula can be used as an estimation. As the amount of heating depends on the initial magnetic field, less heating is reached at 1600 km, where we took the initial magnetic field to be only $B_0=3$ G.

\begin{figure*}[t]
\center
\includegraphics[width=16cm]{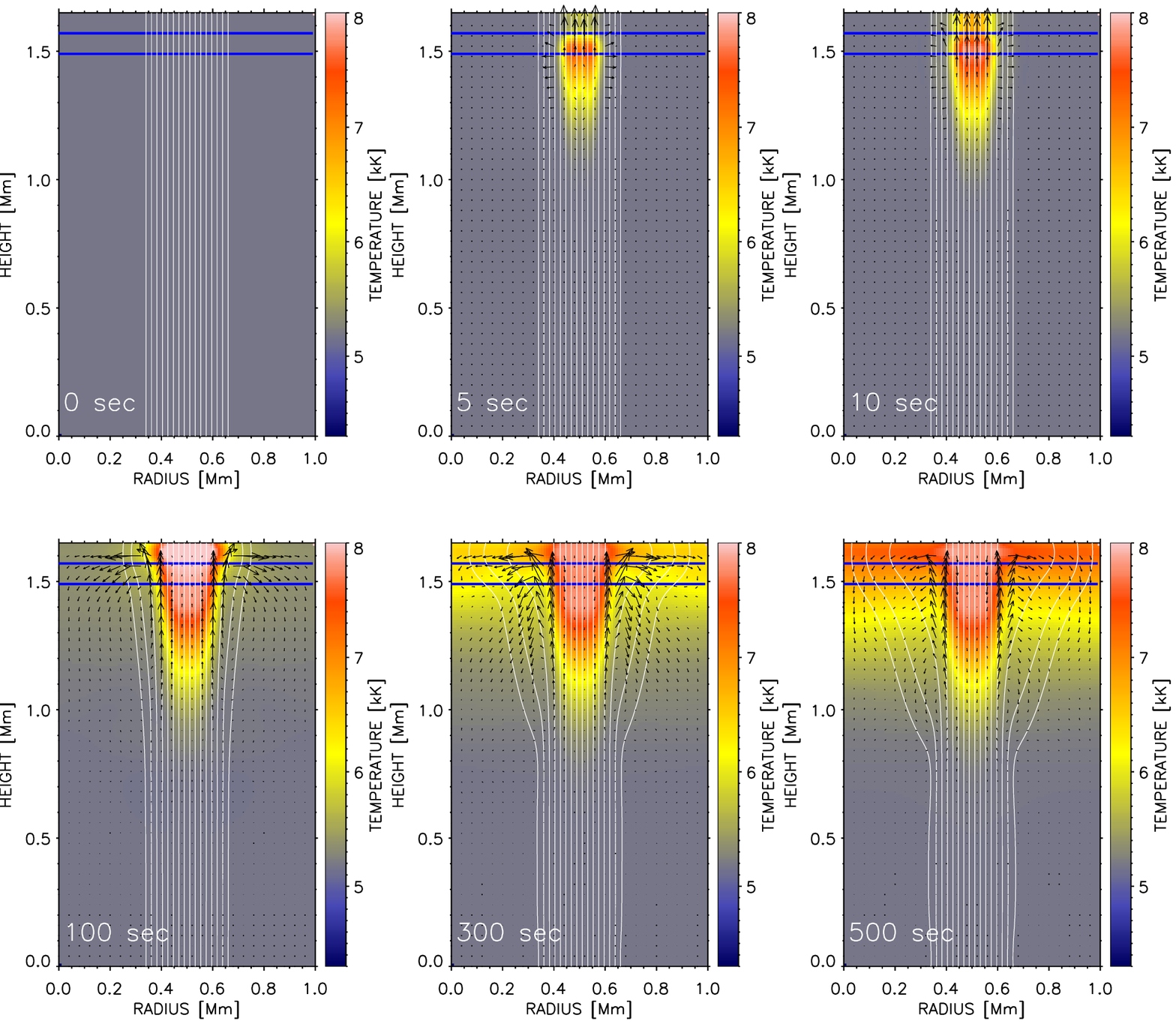}
\caption{{\footnotesize Time evolution of temperature in an (initially) vertical magnetic flux tube described in Sect.~\ref{sect:tuborecto}. The color scale is given on the vertical bars, same for all panels. The background grey color corresponds to a temperature of 5179 K on the color bar}. Vertical white lines are magnetic field lines. Arrows show the velocity field. The initial magnetic field strength is $B_0=300$ G. The time after the start of the simulation is indicated on each panel. Two horizontal blue lines at the top mark the boundaries of the regions where $\eta_A$ is decreased to zero (lower line) and where $\eta_A=0$ (upper line).}
\label{fig:b300evol}
\end{figure*}

\begin{figure*}[t]
\center
\includegraphics[width=16cm]{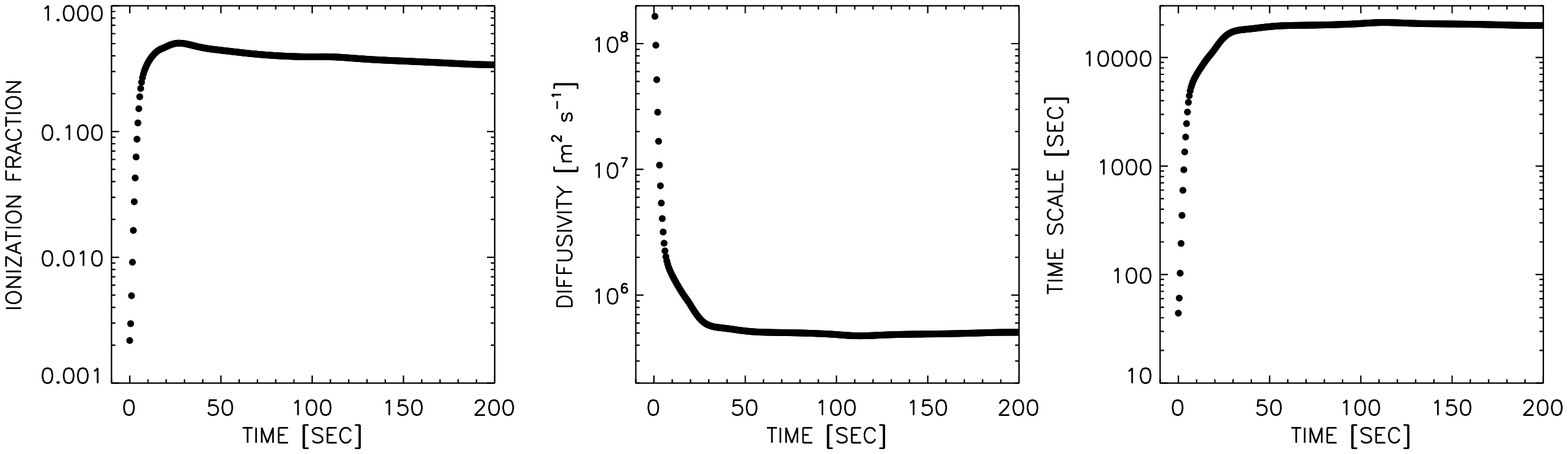}
\caption{{\footnotesize Time variation of the ionization fraction $\rho_e/\rho$ (left); ambipolar diffusion coefficient $\eta_A$ (middle); and characteristic time scale $L^2/\eta_A$ (right) at the height of 1500 km at the flux tube center in the simulation with $B_0=300$ G from  Sect.~\ref{sect:tuborecto}.}}
\label{fig:ionization300}
\end{figure*}

\section{Vertical flux tubes}
\label{sect:tuborecto}

In this section, we describe the results of a similar simulation, but now in a gravitationally stratified isothermal model atmosphere. As initial magnetic field structure, we take the vertical flux tube with $B_z(x)$ given by Eq. (\ref{eq:vtube}). We made five simulations with the value of $B_0$ varying from 300 to 900 G. The atmosphere has initially a constant temperature of $T_0=5179$ K and both pressure and density are stratified with a
scale height of 157 km. Note that now the atmosphere is not in horizontal force balance since $B_z$ is constant with height. The size of the simulation box is 1 Mm by 1.65 Mm with a resolution of 10 km. We take periodic boundary conditions in horizontal direction and closed boundary in both vertical directions. Horizontal motions of the magnetic field lines at the top and bottom boundary are allowed.

As can be appreciated from Fig.~\ref{fig:etas}, the ambipolar diffusion coefficient $\eta_A$ increases exponentially with height. The diffusion term in the energy equation is proportional to $\eta_A$. Thus, we expect the largest heating at the upper part of the simulation domain. To avoid any possible numerical artifact due to the conditions at the upper boundary we artificially decreased $\eta_A$ gradually to zero in the upper 15 grid points (150 km) of the domain. To that aim, we used a third-order polynomial that matches the value of $\eta_A$  and its derivative in one end and reduces down to zero, and zero derivative, at a distance of 7 grid points (70 km). The rest of the points, up to 150 km, have $\eta_A$ equal to zero. This way, we make sure that the heating starts inside the physical domain, not influenced by the upper boundary.

Figure \ref{fig:b300evol} shows the time evolution of the temperature and magnetic field strength in the simulation with initial $B_0=300$ G. The heating starts immediately at the central part of the tube in a few seconds time scale. The appearance of the temperature perturbation at the upper part of the simulation domain goes in agreement with the current distribution in the initial magnetic field configuration and also with the fact that
the ambipolar diffusion coefficient reaches maximum values at 1.5 Mm according to our simulation setup, and then reduced to zero at the topmost layers. It is evident from the figure that, in about 10-seconds time, the temperature increases about 2000 K from its initial value of $T_0=5179$ K at heights around 1.5 Mm, with a much lower variation in those layers where $\eta_A$ is artificially decreased. It must be noted that the perturbation introduced by $\eta_{A}$ also causes plasma motions. At the initial stage of the simulation, these motions are upflows at the central part of the tube. Due to the action of the advective term $\vec{\nabla}(\vec{u} \, e_{\rm tot})$, the velocity motions transport hot material to the upper layers and heat them. This heating is thus a consequence of the transport of hot plasma and not the direct consequence of ambipolar diffusion. The amplitudes of the plasma motion do not exceed 100 m/sec and decrease with time as the tube evolves. The diffusion of the field causes its ``opening'' with height. The field becomes progressively more horizontal with time at heights above 0.8 Mm, which is also a direct consequence of the increase of $\eta_A$ with height. The expansion of the field, together with the outward plasma motions, cause the temperature increase also in the surroundings
of the flux tube, since the temperature perturbation is transported by the advective term of the energy equation. But the heating is the largest in the central part of the tube where the magnetic field is the strongest. Also the time scales of heating inside and outside the flux tube are very different. While the central part of the tube is heated by several kK in a time interval of a few seconds, the surroundings take longer time to evolve and are heated by several hundred K in 5-10 minutes. However, the details of the behavior of the flux tube surroundings are not relevant for our study.

Another visual impression from Fig.~\ref{fig:b300evol} is that the heating at the central part of the tube stops after 100 sec of the evolution. It actually stops sooner, but this time moment is not shown in the figure. The explanation for that is provided in Fig.~\ref{fig:ionization300}. This figure shows the time evolution of the ionization fraction, the ambipolar diffusion coefficient $\eta_A$ and the characteristic time scale given by
the ambipolar diffusion $t \sim L^2/\eta_A$ at the central part of the flux tube at a height of 1.4 Mm. During the first 20 seconds of the evolution, the changes are very significant. The material gets heated and the ionization fraction increases from $10^{-3}$ to 0.5. This produces an immediate decrease of the ambipolar diffusion coefficient by about 3 orders of magnitude. Thus, the time scale also increases abruptly. Initially the heating takes place on about few tens of seconds time scale, but after first 40 seconds the characteristic time becomes about 20000 sec (about 5 hours).
This explains the slowness of the evolution of the temperature at the central part of the tube after the first few tens of seconds.

\begin{figure}[t]
\center
\includegraphics[width=9cm]{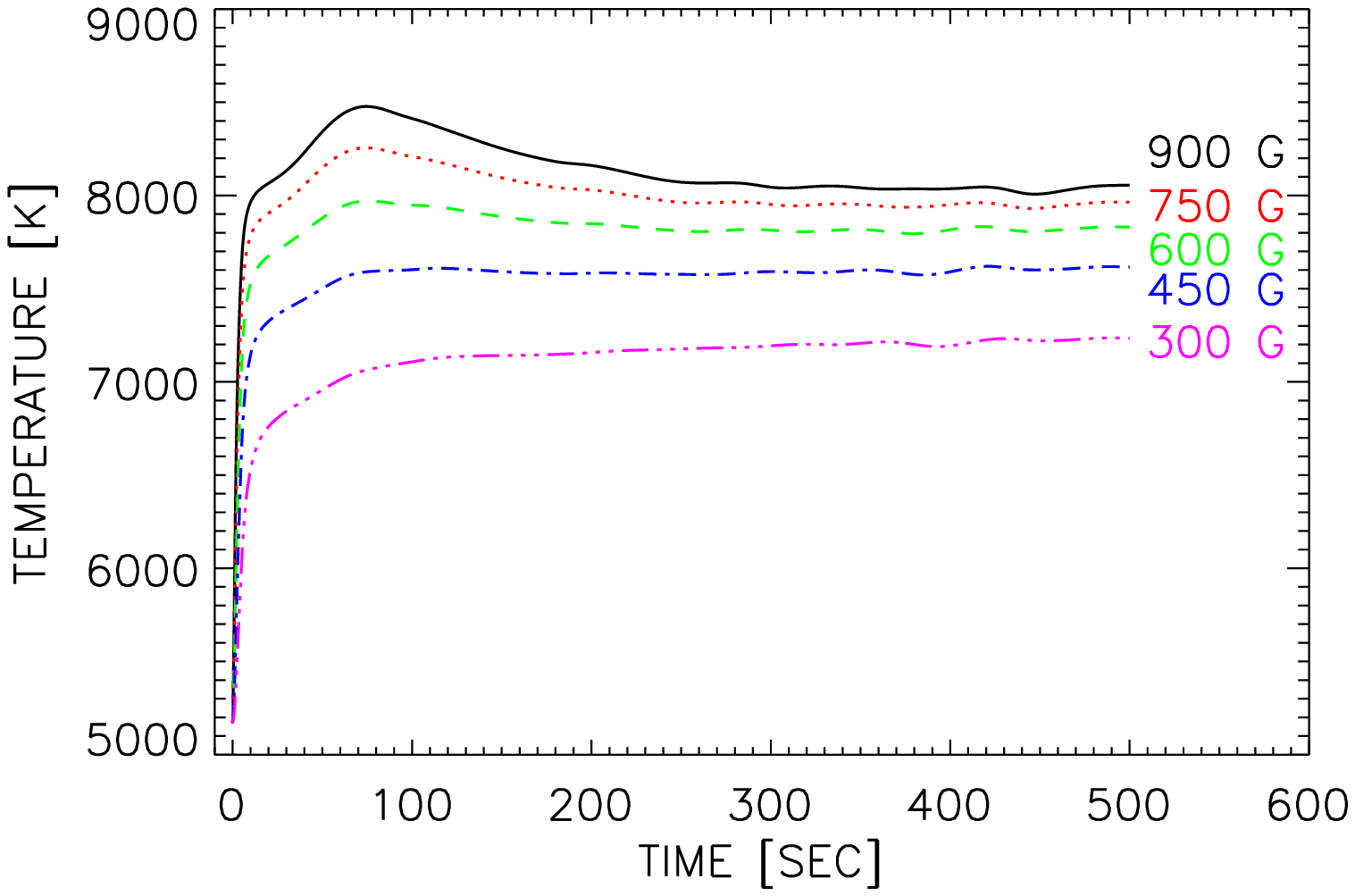}
\caption{{\footnotesize Time variation of the average temperature at heights 1000--1500 km at the center of the flux tube in the simulations with $B_0=300$, 450, 600, 750 and 900 G from Sect.~\ref{sect:tuborecto}. }} \label{fig:timeevol}
\end{figure}

Finally, Figure~\ref{fig:timeevol} gives the temperature variation as a function of time at the center of the flux tube, averaged between 1 and 1.5 Mm heights, for all five simulations with different $B_0$. As expected, since $\eta_A$ depends on the magnetic field, the temperature increase is more significant for the larger fields. $\Delta T$ reaches a maximum of 3.5 kG for $B_0=900$ G at $t=70 $ sec, making  $T=8500$ K. The magnetic field strength has been decreased by only about 3 G. In all cases, the temperature rise is extremely quick during the first few tens of seconds.  It then reaches a stationary stage after $t=100-200$ seconds, depending on the field strength. In the stationary stage, the values of the temperature stay between 7000 and 8000 K. These values are larger than those given by the VAL-C model atmosphere at 1.5 Mm \citep[6400 K, see][]{valc}.

The above simulations have several concerns. Firstly, the vertical flux tube is initially not in horizontal force equilibrium, which might cause an excess of the horizontal plasma motions and other side effects. In this respect, we would like to remind that our code does not solve the full differential equations. After subtracting the equilibrium conditions from Eqs.~\ref{eq:continuity}, \ref{eq:momentum}, \ref{eq:totalenergy}, and \ref{eq:induction}, the resulting differential equations for the evolution of the perturbations are solved. This means that, even if our flux tube is not initially in equilibrium, it cannot evolve if no perturbation is applied to it. In our case, the ambipolar diffusion represents this perturbation, and without it no evolution is present. Thus, the results of our calculation are determined by this parameter and not by the deviations from equilibrium. We can not disregard, though, that the exact evolution (especially of the velocities) may be influenced by the non-perturbed values. These cannot, however, have any influence on the thermal evolution, which is fully determined by the ambipolar diffusion.

The second concern may come from the adopted magnetic field strength, which is, possibly, too large as for a quiet chromospheric region. Measuring magnetic fields in the quiet chromosphere may be challenging \citep[see, e.g.,][]{Manso+Bueno2010} and actual measurements are lacking. In active regions, where the polarization signals are stronger, non-LTE inversions of the infrared \CaII\ triplet and \FeI\ lines have allowed \citet{Socas-Navarro2005, Socas-Navarro2007} to constrain the magnetic field strength to about 1.5 kG above sunspots and about 1 kG in a network region. As far as we are aware, the measurements for the quiet chromosphere are still missing. But, as a reasonable choice, one may assume values of about tens of Gauss.

To address these two concerns, in the next section we describe a simulation done in a more realistic flux tube model in equilibrium, with a lower field strength at chromospheric level, more representative a quiet area.

\begin{figure}[t]
\center
\includegraphics[width=12cm]{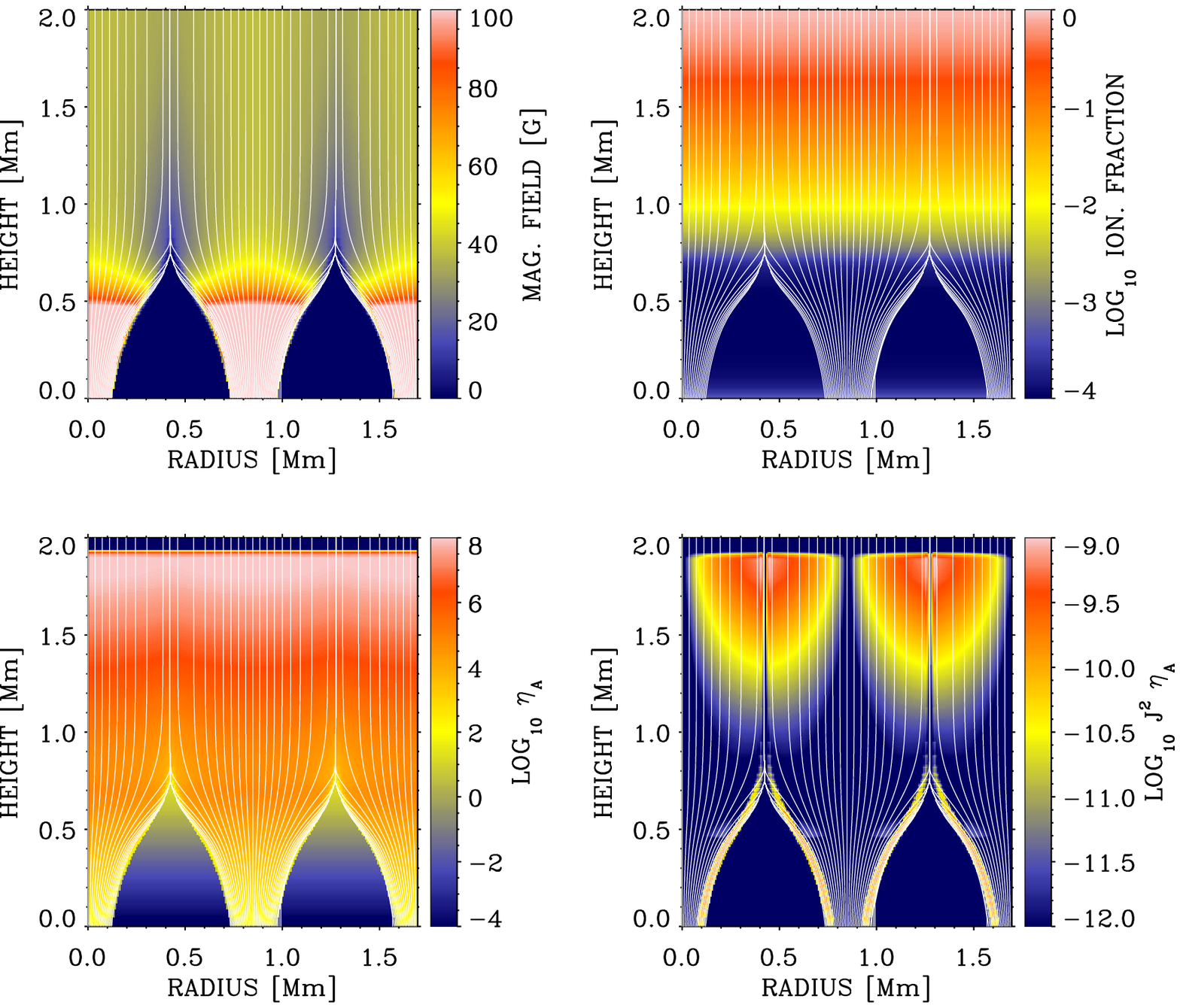}
\caption{{\footnotesize Initial distribution of the magnetic field (upper left); ionization fraction $\rho_e/\rho$ (upper right); ambipolar diffusion coefficient $\eta_A$ (bottom left); and the quantity $J^2 \eta_A$ (bottom right) in the 2nd order flux tube model. Note that $\eta_A$ is decreased artificially to zero at the upper 120 km (15 grid points).}} \label{fig:ft2initial}
\end{figure}

\begin{figure*}[t]
\center
\includegraphics[width=16cm]{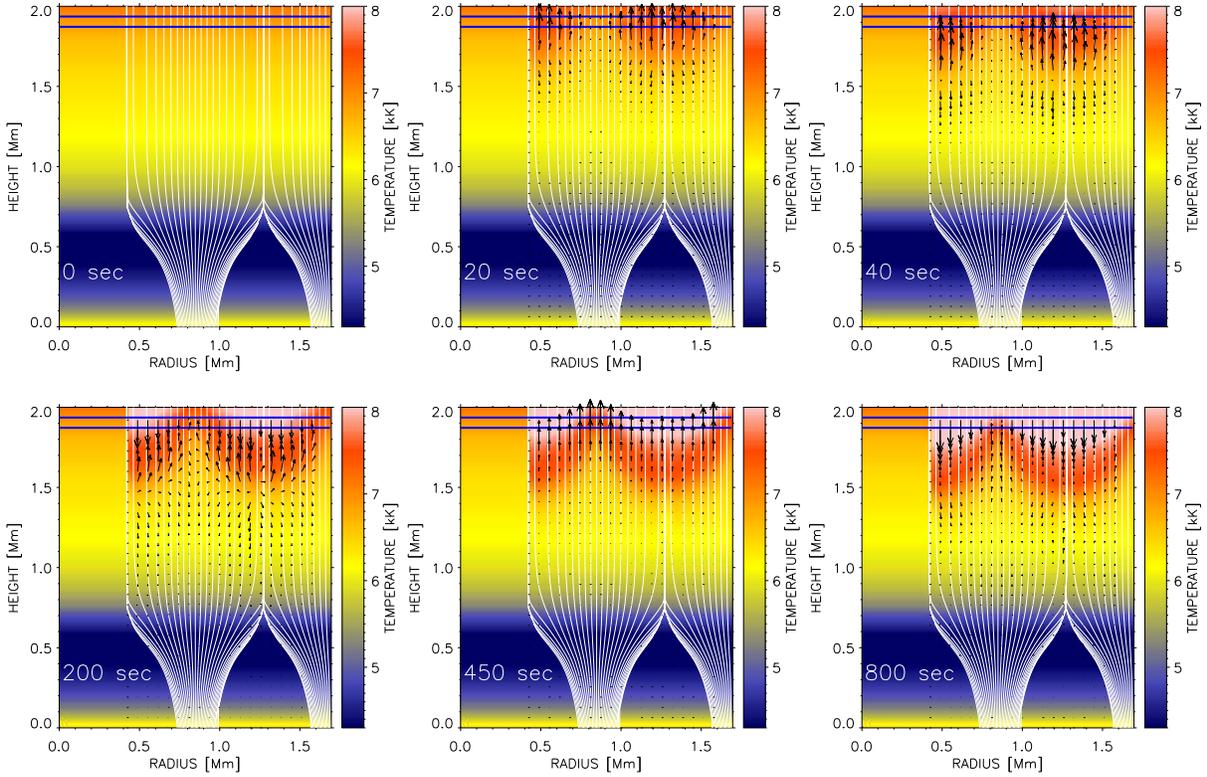}
\caption{{\footnotesize Same as Fig.~\ref{fig:b300evol} but for the 2nd order thin flux tube. For better visual comparison we keep unchanged the temperature structure at the left hand side of the domain (from 0 to 0.42 Mm)}, though the variations are present in the simulations. }
\label{fig:ft2evol}
\end{figure*}

\begin{figure*}[t]
\center
\includegraphics[width=16cm]{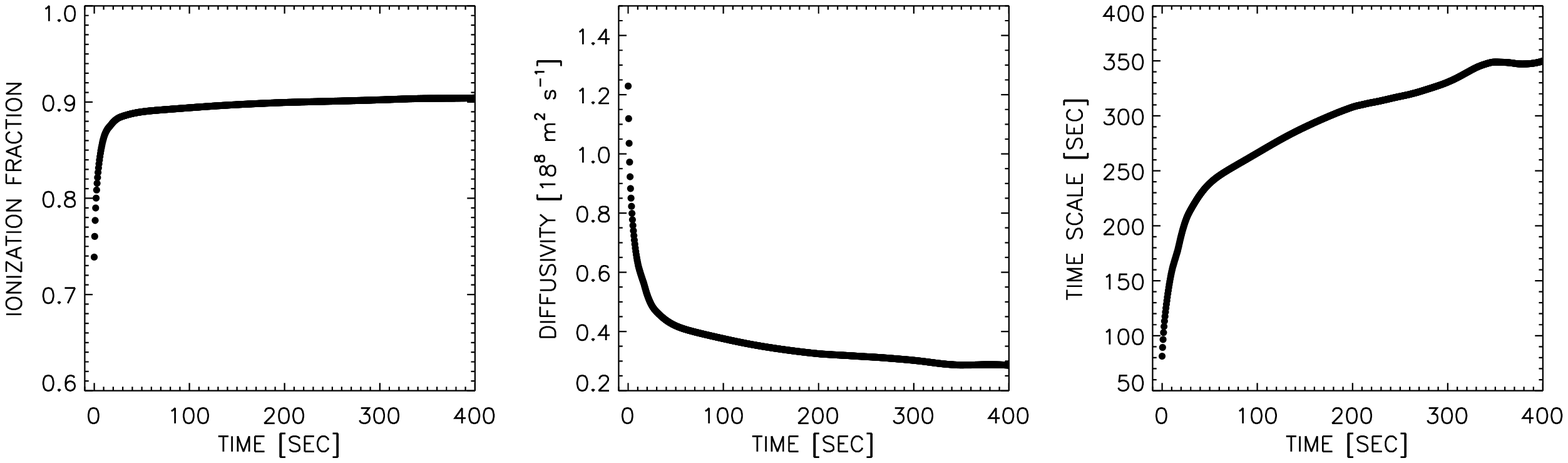}
\caption{{\footnotesize Same as Fig.~\ref{fig:ionization300} but for the 2nd order flux tube model at horizontal distance 0.6 Mm and height 1900 km.}} \label{fig:ft2ionization}
\end{figure*}

\begin{figure}[t]
\center
\includegraphics[width=9cm]{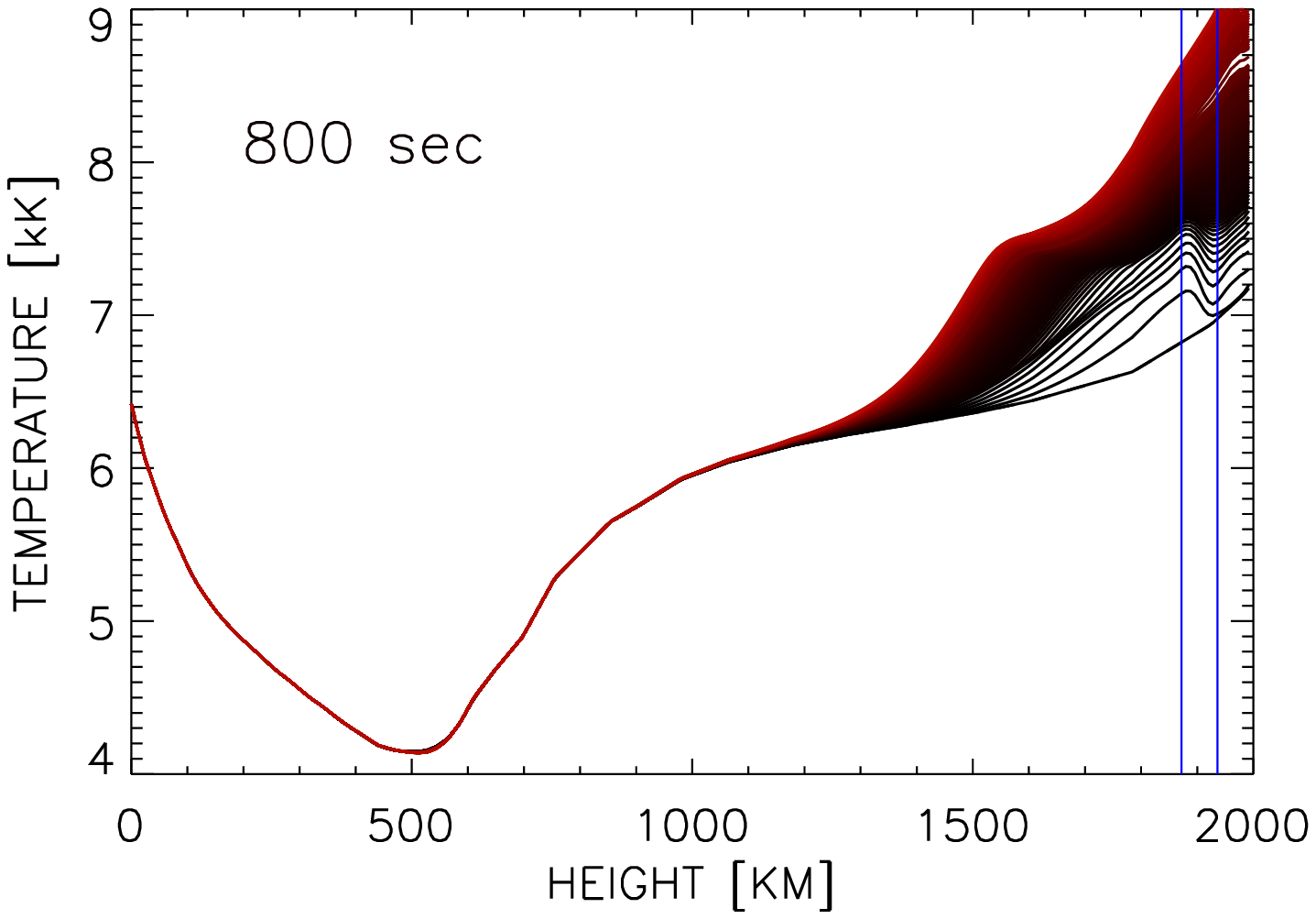}
\caption{{\footnotesize Temperature as a function of height at horizontal distance 0.6 Mm of the 2nd order flux tube from Fig.~\ref{fig:ft2evol}. Different lines are separated 20 sec in time with progressively more red colors indicating larger times till 800 sec since the start of the simulation. Note that the evolution gets slower with time. Two horizontal blue lines mark the boundaries of the regions where $\eta_A$ is decreased to zero (1880 km) and where $\eta_A=0$ (1940 km).}} \label{fig:ft2temp}
\end{figure}

\section{Second-order thin flux tube}

To simulate non-active chromospheric conditions, we used the 2nd-order thin magnetic flux tube as initial model atmosphere \citep{Pneuman+etal1986, Khomenko+etal2008}. Previously, the same flux tube model was used for simulations of wave propagation from the photosphere to the chromosphere and it is described in the corresponding papers \citep{Khomenko+etal2008, Khomenko+etal2008b}. The model represents a series of flux tubes that merge after some height in the chromosphere, preventing them from excessive opening with height. Thus, the magnetic field configuration is non-potential and there are currents in the system. Figure \ref{fig:ft2initial} illustrates the initial distribution of some parameters of the flux tube model. The magnetic field strength (Fig.~\ref{fig:ft2initial}, upper left) has both vertical and horizontal gradients. The colors of the figure are saturated at the bottom part of the structure to highlight the variations of the magnetic field in the chromosphere. The field is stronger at the central part of the
tube, with about 37 G from 1 Mm upwards. The ionization fraction defined as $\rho_e/\rho$ (Fig.~\ref{fig:ft2initial}, upper right) varies in agreement with the temperature distribution of the VAL-C model atmosphere. It has its lowest values in the photosphere, dropping down to $10^{-4}$ and reaching about 0.7 in the upper part of the domain, at 1.5 Mm. The ambipolar diffusion coefficient (Fig.~\ref{fig:ft2initial}, bottom left) changes from $10^2$ in the photosphere up to $10^8$ in the chromosphere inside the flux tube (note its artificial decrease in the upper 15 grid points, 120 km, done for the same reasons as in the simulation shown in the previous section). Outside the flux tube, its value is very small, as the magnetic field strength outside the tube is negligible (we set magnetic field outside the flux tube equal to some small value for reasons of numerical stability of the simulations). The most interesting quantity is the one shown at the bottom right panel of Fig.~\ref{fig:ft2initial}. It gives a proxy of the diffusion term calculated as the square root of the current $J^2$ multiplied by the ambipolar diffusion coefficient (see Eq.~\ref{eq:intenergy}). As the variations of the magnetic field are stronger near the borders of each individual tube, the current is also stronger there. The absolute value of $J^2$ decreases according to the drop of the magnetic field strength with height, while $\eta_A$ exponentially increases. When multiplied one by the other, the quantity $J^2\eta_A$ has largest values at the upper part of the domain above $1-1.3$ Mm, near the borders of the tubes. This term is responsible for the heating, therefore we expect to observe the largest heating of the flux tube atmosphere at the locations where $J^2\eta_A$ is important.

Figure \ref{fig:ft2evol} gives several snapshots of the evolution of temperature in the flux tube after the diffusion terms were introduced as perturbation. Same as before, the boundary conditions are taken periodic in the horizontal direction (equivalent to an infinite series of flux tubes); the domain is closed at the top and bottom boundaries; and $\eta_A$ is decreased smoothly to zero at the upper 100 km of the simulation domain. Except for velocity variations (unimportant for the conclusions of the present study), closed boundary conditions do not affect the variations of any other quantity in our simulations.

Similar to the case of vertical flux tubes (described in Sect.~\ref{sect:tuborecto}), the heating starts at the upper part of the domain, close to the tube borders.  This behavior is expected because the term responsible for the heating ($\sim J^2\eta_A$, see Fig. \ref{fig:ft2initial}) is orders of magnitude larger at these locations. The upper 120 km of the atmosphere, with artificially reduced $\eta_A$, are not heated during the first few seconds, although the heat is transported there by the advective term in the energy equation from the material just heated below. The heating gradually extends beyond the tube borders and extends down into the atmosphere. The temperature at 1900 km reaches 8800 K after 800 sec of the simulation, which is about 2000 K above its initial value. The time scales of the heating are longer compared to the case of the vertical tubes with larger field strength. There is still no saturation after 800 sec of the simulation and the temperature is increasing slowly with time.

The velocity field developed in this simulation is different from the case of vertical flux tubes (Sect.~\ref{sect:tuborecto}). The motion is mostly vertical and periodic in time, changing from upflows to downflows. Such motion is a direct consequence of the closed upper and lower boundary conditions and the periodicity of our magnetic structure (series of flux tubes in horizontal direction with mostly vertical field in the chromosphere). There is also no expansion of the field with height. The expansion is not possible as the tubes are located one next to the other.

Temporal variations of the ionization fraction, ambipolar diffusion coefficient and the parameter $L^2/\eta_A$ are shown in Fig.~\ref{fig:ft2ionization}. The values are taken at 1900 km close to the border of the tube at horizontal location $X=0.6$ Mm. The initial ionization fraction at this height is already rather high. During the first 50 seconds of the simulation, it increases from 0.7 to about 0.9. Accordingly, the ambipolar diffusion coefficient drops by about an order of magnitude after the first 50 seconds. Later, the evolution slows down and the further decrease of $\eta_A$ is more gradual. The change of the characteristic time scale is shown in the right panel of Fig.~\ref{fig:ft2ionization}. The time scale increases from 50 seconds (initially) up to 300--400 seconds (after 400 sec of the simulation).

In the case of the vertical tubes (Sect.~\ref{sect:tuborecto}), the variations of $t \sim L^2/\eta_A$ were stronger and the values of $t$ at the later stages of the evolution were much larger. The difference appears because, firstly, the initial magnetic field strength of the vertical tubes was much larger, causing a rapid heating and ionization within the first 20 seconds of the simulation. Secondly, we considered lower heights up to 1500 km.
The value of $\eta_A$ changes more than one order of magnitude between 1500 and 2000 km (Fig.~\ref{fig:etas}) and so does the time scale. Thus, while the evolution gets slower in time in the simulation with a 2nd order flux tube, it is not as slow as in the case of the vertical tubes. After the initial phase of the rapid evolution, the temperature in our 2nd order flux tube continues increasing on scales about 5--6 minutes. Obviously, in a more realistic modeling, this constant energy input is expected to be balanced by some dissipation mechanism. Otherwise, the simulation would eventually become unstable. However, here we are interested in understanding and evaluating the action of the heating (not cooling) term. For that reason, we considered a relatively simple situation that allows us to better constrain the physics. Clearly, the final judgment about the heating efficiency of the proposed mechanism will require the inclusion of radiative energy exchange, a key mechanism for the chromospheric energy losses \citep{Athay1966}.

Finally, Figure \ref{fig:ft2temp} gives the temperature as a function of height at the horizontal location $X=0.6$ Mm, close to the tube border, at different time moments. The atmosphere gets heated above 1100 km and the amount of temperature increase is larger at larger heights, except in the uppermost layers at the beginning of the evolution, where the ambipolar diffusion does not operate. The magnetic field strength has decreased on average by about 0.01 G at heights 1100--1900 km and $X=0.6$. Thus, we have dissipated only about 0.03\% of the existing magnetic field. Fig.~\ref{fig:ft2temp} can be compared with Fig.~\ref{fig:dt}, where we calculated the temperature increase $\Delta T$ assuming the same model atmosphere, but with 1\% of the existing magnetic field at each height being dissipated. Given that we have dissipated 30 time less magnetic field than it was assumed for the calculations of $\Delta T$ in Fig.~\ref{fig:dt}, the results of our numerical simulations are in order of magnitude agreement with the simple estimation by Eq.~\ref{eq:formula}.

\section{Conclusions}

In this work, we have investigated the heating of the solar magnetized chromospheric plasma produced by current dissipation (Joule heating). The current dissipation is enhanced by orders of magnitude due to the presence of a significant amount of neutral atoms in the chromospheric plasma not entirely coupled by collisions. The net relative motion between the ionized and neutral plasma components results in an additional diffusion term
(ambipolar or neutral diffusion). The value of the ambipolar diffusion coefficient depends on the fraction of neutral atoms and on the magnetic field strength and can be as much as five orders of magnitude larger in the chromosphere than the classical ohmic diffusion coefficient. This implies evolution time scales of about 10--100 seconds. Thus, all the processes related to the ambipolar diffusion, including the plasma heating, happen very quickly.

We have performed analytical estimates and numerical simulations in different models of the amount of plasma heating due to diffusion by neutral particles. The simulations covered a variety of scenarios, from a simple atmosphere without gravity to a more complex flux tube model, resembling a chromospheric quiet network area. All our calculations show that the lower chromosphere can be  heated on temporal scales of $10^2$ seconds, giving rise to a temperature increase of several kK. The only requirements for the heating is the existence of a non force-free
magnetic field. We obtain that the temperature increase is significant above a height of 1 Mm, depending on the strength of the magnetic field.

From those results, we conclude that current dissipation enhanced by the action of ambipolar diffusion is an important process that is able to provide a significant energy input into the chromosphere. We suggest that this mechanism should be included into the future models of chromospheric heating. Here we would like to stress that, apart from possible dissipation mechanisms, the action of the ambipolar term in the energy equation would stop in two cases. Firstly, this will take place if the temperature increase is so significant that all atoms become ionized leading to $\eta_A=0$. Obviously, this is not the case of the chromosphere which is observed to be, at least, partially neutral. Note that our calculations indicate that the time scales of heating may become very long after a fast initial temperature increase (see Figures \ref{fig:ionization300} and \ref{fig:ft2ionization}). This would mean that the actual complete ionization is never reached in a reasonable amount of time. Secondly, the heating due to ambipolar diffusion will stop if the magnetic field configuration becomes force-free, meaning $\vec{J} \parallel \vec{B}$, i.e. $\vec{J}_{\perp}=0$. This might lead to different temperatures and ionization fractions for different starting magnetic field configurations and will need further detailed investigation. Finally, the action of the ambipolar diffusion term has to be balanced by a dissipation mechanism. It is well known since the 60's that the chromosphere is cooled by radiative losses \citep{Athay1966}. Whether the proposed mechanism is able to balance these losses, and whether it is able to provide sufficient energy to support the chromospheric high temperature, has to be investigated in future realistic 3D MHD chromospheric simulations, including a detailed radiative transfer and all complex chromospheric physics.

Here, we would just like to give hints that the temporal scales of the heating are defined by the  magnetic field strength via the ambipolar diffusion coefficient $\eta_A$ (Eq.~\ref{eq:etaa}). For the stronger fields considered by us ($B=300-900$ G), the heating happens very quickly, achieving a temperature increase of 2--3.5 kK in a 1 minute time. For the weaker field strengths (30--40 G), expected in the quiet chromosphere, the plasma is heated by 2 kK in a 10 minutes time. These time scales are comparable to the scales of radiative cooling of the plasma \citep[a simple Newton formula gives characteristic cooling times at this height of the order of $10^2-10^3$ seconds, see][figure 3]{Mihalas+Toomre1982}. This argument may be used to suggest that one process might, in fact, balance the other.

The radiative damping would tend to decrease the temperature in the chromosphere. This, in turn, would produce a decrease of the ionization fraction, a subsequent increase of $\eta_A$ and a decrease of the heating time scale (see Figs.~\ref{fig:ionization300} and \ref{fig:ft2ionization}). The neutral diffusion term would be put in action again, leading to a new temperature enhancement. The detailed temperature balance would depend on the details of non-LTE chromospheric radiative transfer and on the scales of the non-LTE ionization equilibrium.

The present study has several limitations. The calculation of the ionization balance is done in LTE conditions, assuming instantaneous ionization $-$ recombination. Several recent studies have shown that the {ionization-recombination} of (at least) hydrogen and calcium in the solar chromosphere does not follow LTE relations or instantaneous statistical equilibrium \citep{Leenaarts2006, Wedemeyer2011}. In addition, we have discussed the energy losses by radiative damping, but have not included them in our simulations. Thus, in the simulations shown in this paper, the temperature would ever increase until the chromosphere is (unrealistically) completely ionized or until the magnetic field configuration becomes force-free. This, however, was done on purpose since we wish here to investigate the action of a newly proposed mechanism in simple situations, to better understand its implications into the general picture. In our future studies, we plan to include the effects of radiative damping in our calculations.

\acknowledgments This work is partially supported by the Spanish Ministry of Science through projects AYA2010-18029 and AYA2011-24808. This work contributes to the deliverables identified in FP7 European Research Council grant agreement 277829, ``Magnetic connectivity through the Solar Partially Ionized Atmosphere'', whose PI is  E. Khomenko (Milestone 4 and contribution toward Milestones 1).


\end{document}